\documentclass[%
 reprint,
 amsmath,amssymb,
 aps,
]{revtex4-2}

\usepackage{graphicx}
\usepackage{dcolumn}
\usepackage{bm}
\usepackage{xcolor}
\usepackage[normalem]{ulem}
\usepackage{float}

\begin{document}

\preprint{APS/123-QED}

\title{Quantum Optical Electron Pulse Shaper}

\author{Neli Laštovičková Streshkova}
\email{neli.streshkova@matfyz.cuni.cz}
\affiliation{Charles University, Faculty of Mathematics and Physics, Ke Karlovu 3, 121 16 Prague 2}
\author{Martin Kozák}%
\email{m.kozak@matfyz.cuni.cz}
\affiliation{Charles University, Faculty of Mathematics and Physics, Ke Karlovu 3, 121 16 Prague 2}

\date{\today}

\begin{abstract}
Coherent control of ultrafast quantum phenomena benefits from pulse-shaping capabilities allowing to modulate the envelope and instantaneous phase of optical fields on femtosecond time scales. While such control is available for optical fields, an analogy of a pulse shaper for freely propagating electrons is lacking. In this study, we theoretically demonstrate a method that enables near arbitrary light-based shaping of electron wave packets in the time domain. The method is based on the quantum phase modulation of electron waves by coherent light with time-dependent frequency leading to generation of spectrally separated electron energy side bands with shaped time-energy profiles and envelopes. Our results show that few femtosecond time durations can be achieved without additional spectral broadening of the electron wave packet, allowing one to reach the combination of high time, spatial, and spectral resolutions in ultrafast imaging and diffraction experiments with pulsed electron beams.
\end{abstract}

\maketitle

\textit{Introduction---}Pulsed electron sources have been developed with the aim of extending the exceptional spatial resolution of electron microscopes to the time domain. Ultrafast electron microscopy and diffraction evolved in complex tools enabling observation of dynamical phenomena occurring on nanoscale, such as
structural and phase transitions \cite{Kim2023, Danz2021}, magnetic \cite{Berruto2018, Shimojima2023} and charge carrier dynamics \cite{Zandi2020, Perez2024, Sun2015}, atomic motion \cite{Lee2022} and plasmonic \cite{Liu2021} and phonon \cite{Gnabasik2022, Zhang2019} dynamics. 

Pulsed electron beams for ultrafast electron microscopy are routinely generated via laser-stimulated photoemission. The energy bandwidth of the emitted electron wave packet is determined by the geometry and material properties of the emitter, the applied electric field, and the spectrum of photoemission light. The nonzero energy spread combined with the dispersive propagation of electrons lead to elongation of the generated pulses during their acceleration and propagation to the specimen \cite{BAUM2013}. As a result, the shortest pulse duration of electrons achieved in electron microscopes without additional compression is on the order of few hundreds of femtoseconds \cite{Feist2017, Kozak2018JAP, Moriova2025}. Such time resolution allows one to study picosecond processes, but the time scales associated, for example, with coherent lattice vibrations or coherent electron dynamics are not directly accessible without additional compression of the electrons. 

Electron pulse compression schemes are based on time-correlated spectral broadening through interaction with time-dependent electromagnetic forces and subsequent compression due to dispersive propagation of the broadened electron pulse \cite{Hilbert2009, Kaminer2023, Kealhofer2016}. Such manipulation can occur in either the classical or the quantum-coherent regime and has been utilized for the compression of electron pulses to a few tens of femtoseconds durations when utilizing radio-frequency fields \cite{Gliserin2015} or even attosecond durations when using optical fields \cite{Feist2015,Priebe2017,Kozak2018_Phys,Kozak2018,Morimoto2018,Vanacore2018,Kuttruff2024}. However, experiments requiring both high spectral and temporal resolutions would greatly benefit from a technique that enables compression of electron pulses in the time domain while keeping the energy bandwith of the electrons reasonably narrow to maintain spatial and spectral resolution for imaging and diffraction experiments.

Although the phase and envelope of ultrashort optical pulses can be manipulated in the time domain by various shaping methods \cite{Weiner2000,weiner2011ultrafast}, such complex control has not been available for pulsed electron beams. Optical pulse shaping techniques are typically based on separate control of the phase and amplitude of different spectral components of the pulse \cite{weiner2011ultrafast}. Their coherent superposition then determines the time profile of the pulse envelope. However, ultrashort electron pulses differ strongly from their optical counterparts due to the fact that their coherence time is much shorter than the pulse duration \cite{Baum2021}. In contrast to fully coherent optical waves generated by lasers, the electron wave packets from time-separated regions within the same pulse do not coherently add up to shape the envelope of the pulse. While application of phase-modulated laser pulses to control spectral content of X-ray photons emitted by Thomson scattering has been proposed \cite{Shadwick2013, Krafft2014}, complex temporal shaping of the quantum state of the electrons by temporally modulated optical pulses has not been considered yet.\\

\textit{Electron Pulse Shaping Principle---}In this study, we propose a quantum-optical electron pulse shaping concept based on the interaction of a partially coherent electron pulse with shaped optical fields of time-varying frequency.

The stimulated inelastic interaction between a periodic vector potential of quasi-monochromatic light and the electron wave function can be understood in a semi-classical and nonrecoil approximations as a periodic phase modulation of the electron wave function \cite{Gracia2010_Rev}. 
Efficient modulation of electron energy can be enabled using plasmonic nanostructures \cite{Feist2015, Piazza2015, Gracia2016, Priebe2017, Harvey2020, Talebi2020, Kozak2018JAP}, metallic or dielectric membranes \cite{Vanacore2018, Vanacore2018a, Andrea2023, Feist2020, Morimoto2018}, evanescent optical fields \cite{Breuer2013, Peralta2013, Kozak2017, Dahan2020_Nature, Henke2021, Kfir2020}, or ponderomotive coupling with multiple optical beams in vacuum \cite{Kozak2018_Phys, Kozak2018, Tsarev2023}. The final wave function after the interaction can be described as a superposition of discrete energy sidebands separated by the energies of the interacting photons $\hbar\omega$ \cite{Reinhardt2020}, where $\omega$ typically represents a time-independent frequency of the optical field.

Our proposal is to generalize this principle to the case of optical fields with time-dependent photon energy $\hbar\omega(\tau)$, allowing to directly control the immediate sideband separation in phase space. Here we assume that the optical frequency $\omega(\tau)$ changes slowly compared to the electron coherence time. See Supplemental Materials (SM) \cite{SM} for detailed derivation of the effect of the light field with time-dependent frequency on the electrons.

\begin{figure}
    \centering
    \includegraphics[width=\linewidth]{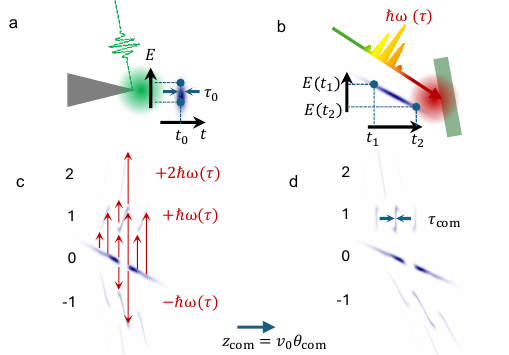}
   
    \caption{ (a) An electron pulse with finite spectral width is generated via short pulse emission from the tip. (b) The electron pulse acquires a positive chirp through dispersive propagation. At the interaction site the electron pulse traverses a dielectric membrane e.g., illuminated by an optical pulse with modulated instantaneous photon energy $\hbar \omega(\tau)$ and intensity envelope $I(\tau)$. (c) The absorptions (emission) of photon quanta result in the generation of tailored energy sidebands. (d) After propagating for a certain distance, the quesi-probability side bands are reshaped.
    }
    \label{fig:Scheme}
\end{figure}

The electron pulse is generated by photoemission from a photocathode [Fig. \ref{fig:Scheme} (a)]. We assume that the duration of an electron pulse immediately after photoemission is determined by the time envelope of the optical pulse incident on the cathode in the electron source \cite{BAUM2013}. In contrast, the electron coherence time depends on the photoemission conditions. Without additional monochromatization, the typical electron energy spread in an ultrafast transmission electron microscope equipped with a field emission source is $\delta E=$0.3-0.7 eV \cite{Feist2017,Schröder_2024} with the associated coherence time of electrons $\tau_\text{coh}=\hbar /\delta E\approx$2-5 fs \cite{BAUM2013, Baum2021}.
The electron pulse is then accelerated by a static electric field, and afterwards propagates freely to the pulse-shaping plane, while both of the processes contribute to temporal elongation and chirping of the electron wavepacket. We consider an electron wavepacket with less than one electron per pulse, thus we neglect possible space charge effect on the electron model properties and propagation.

In the pulse shaping plane, the electron pulse interacts with a shaped laser pulse at a dielectric membrane [Fig. \ref{fig:Scheme} (b)], ensuring efficient coupling of electrons and photons \cite{Morimoto2018a}. The electric field of the laser pulse generated in an optical pulse shaper \cite{Weiner2000,weiner2011ultrafast} can be described using an envelope and a time-dependent frequency $\omega(\tau)$. Due to the coherent phase modulation of the electron wave function at the instantaneous frequency of the shaped light pulse, the separation of the generated electron energy sidebands $\hbar\omega(\tau)$ becomes a function of time $\tau$ [Fig. \ref{fig:Scheme} (c)]. Moreover, their instantaneous populations can be controlled by the envelope of the optical pulse. These two virtually independent quantities give us a tool for complex and general shaping of the electron wave in the longitudinal phase space.

Combined with subsequent dispersive propagation between the pulse shaping plane and a sample, the electron pulse can be structured in time domain leading to a single compressed pulse or a periodic train of compressed pulses in one of the electron energy sidebands [Fig. \ref{fig:Scheme} (d)]. Thanks to the sufficient separation between the sidebands, the shaped sideband can be isolated from the rest of the electron distribution by spectral filtering.\\

\textit{Results---}The pulse shaping method is examined in two model situations for lower kinetic energy electrons (5 keV) and higher kinetic energy electrons (30 keV). In each special case we aim to reduce the duration of the pulse or pulses in the specimen plane to few fs.

In the following section, we show the electron pulse in energy-time phase space represented by a non-negative spectrogram $S(E,\tau)$ defined as a convolution of the Wigner function $W(E,\tau)$ of the electron wave-packet with a Gaussian kernel
\begin{equation}
    S(E,\tau) = W(E,\tau)*\exp{[-\tau^2/\tau_s^2]},
 \end{equation}
where $\tau_s \approx 5\,\mathrm{fs}$. By definition the Wigner function is allowed to exhibit negative values. We opt for spectrogram representation, because it allows us to better visualize the energy band populations and smooths out the fast oscillating coherences, which occur between the bands \cite{Me2024}. Detailed description of the theory and numerical simulations can be found in SM \cite{SM}.\\

\textit{Chirp Inversion Induced Pulse Compression---} A special case of pulse shaping is temporal compression of electron pulses induced by chirp inversion in a single energy sideband. We use as an example low energetic electrons with central kinetic energy of $5\,\mathrm{keV}$. Low-energy electrons, in general, undergo significant temporal elongation due to dispersion already after short distances, reaching durations several times longer than the initial length of the pulse. Also, the compression of the low energetic electrons can be conveniently achieved at relatively short propagation distances.

\begin{figure}
    \centering
    \includegraphics[width=1\linewidth]{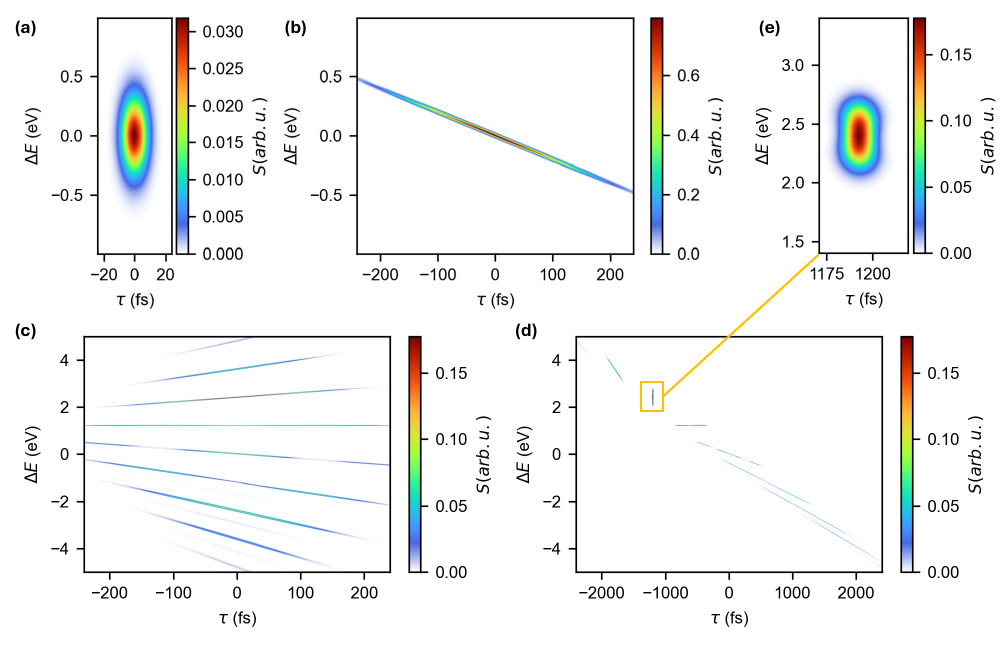}
    \caption{Spectrogram representation of the electron pulse (a) dispersion-free, with total pulse duration of 9.9 FWHM, (b) with group delay dispersion $\alpha$ and total duration of 250 fs FWHM, (c) the electron side-bands after interaction with the optical field, (d) after propagation, (e) close-up of the compressed sideband spectrogram. We note that in the spectrogram representation the sidebands are elongated in the $\tau$-direction due to the convolution with Gaussian kernel.
    }
    \label{fig:Compression}
\end{figure}

After photoemission, the electron wavepacket has an envelope with duration $\tau_0 = (\tau_{\text{coh}}^2 + \tau_{\text{pe}}^2)^{1/2}= 9.9\,\mathrm{fs} $, with coherence time $\tau_\text{coh} = 3.65\,\mathrm{fs}$, photoemission pulse duration $\tau_\text{pe} \approx 9.30 \,\mathrm{fs}$ and spectral width $\delta E = 0.5\,\mathrm{eV}$ [Fig. \ref{fig:Compression} (a)]. We consider acceleration to $5\,\mathrm{keV}$ (central group velocity $v_0 = 0.14c$) and dispersive propagation, resulting in elongation to total duration of $\tau_\text{el}=250\,\mathrm{fs}$, due to second order group delay dispersion (GDD) term $\alpha = 328\,\mathrm{fs^2} $ [Fig. \ref{fig:Compression} (b)].

The spectral domain representation of the optical pulse used for chirp inversion is explicitly
\begin{equation}
    \tilde{\mathcal{E}}(\omega)\propto\exp\left[-\frac{(\omega - \omega_0)^2}{2\sigma^2_\omega} - i\frac{\varphi''(\omega - \omega_0)^2}{2}\right],
 \end{equation}
where $\omega_{0} = 1.8 \,\mathrm{fs^{-1}}$ is the central photon frequency, the spectral width is matched to that of the electron wave packet $\sigma_\omega = \delta E/\hbar\sqrt{4\ln(2)}$, the energy width, and the opposite GDD coefficient is used $\varphi'' = -\alpha$.

The spectrogram [Fig. 2 (c)] of the electron pulse after the interaction with light consists of the ladder of energy bands spaced by integer multiples of the instantaneous photon energy. We consider constant coupling strength of the interaction $g = 1.75$ (see \cite{SM} for definition), so about 20\% of the total electron probability distribution is transferred to the second-order sideband, where the reversal of the initial chirp is achieved.

We can derive the required propagation time $\theta_{\text{com}}$ for the second side-band in the electron probability density to compress. We consider that the photon energy is small compared to the central electron energy $E_0\gg\hbar\omega$ and the group velocity $v(E)$ in the vicinity of $v_0$ varies linearly with the energy $[v(E)-v_0] \propto E$. With central group velocity of the second sideband $v_{2\hbar\omega_0}=v(E_0+2\hbar\omega_0)$, the electron pulse needs a propagation distance of $z_{\text{com}}=\frac{m_e v_{2\hbar\omega}^2\alpha}{\hbar}\approx21\,\mathrm{cm}$. We label this position in  $z$ as the compression point. The second sideband reaches the initial duration of the pulse (numerically evaluated $\approx9.9\,\mathrm{fs}$) reversing the elongation while maintaining the initial spectral width. [Fig. \ref{fig:Compression} (d), (e)]. \\

\textit{Nonlinear Electron Chirp Correction---}To demonstrate the capabilities of electron pulse shaping for correction of higher order dispersion, we add a third order dispersion term (TOD)  $\beta = 256\,\mathrm{fs^3}$ to the electron pulse with the same parameters as used in the calculations presented in Fig. 2. In this case the electron pulse phase space representation is curved and more importantly, correction with linearly chirped light is not sufficient.

\begin{figure}
    \includegraphics[width=\linewidth]{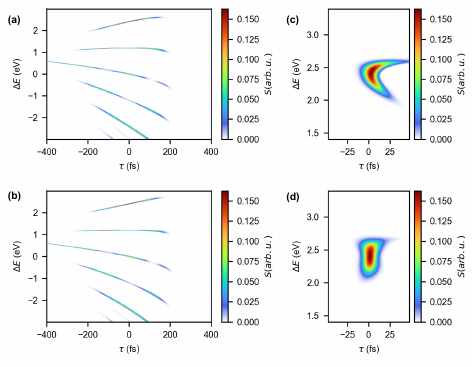}
    \caption{
     Spectrogram of the electron pulse with 242 fs FWHM with linear and nonlinear chirp (a) after interaction with linearly chirped light (GDD only) (b) after interaction with light with GDD and TGD (c) GDD compensated after propagation for 21 cm, compression to 13 fs FWHM (d) after propagation for 21, GDD and TGD compensated, compressed down to 11 fs.
    }
    \label{fig:03}
\end{figure}

In Fig. \ref{fig:03} we show the spectrogram of an electron pulse with a combination of linear and nonlinear chirp. In Fig. \ref{fig:03}(a), optimally linearly chirped light is used for the correction. After propagation for 21 cm it is noticeable, that the sideband generated with linearly chirped light cannot compress ideally, reaching 13 fs FWHM duration [Fig. \ref{fig:03}(c)] due to the leftover curvature of the band.

To compensate for $\beta$, we need to introduce TOD $\varphi'''$ into the optical pulse as well:
\begin{equation}
    \tilde{\mathcal{E}}(\omega)\propto e^{\left[-\frac{(\omega - \omega_0)^2}{2\sigma^2_\omega} - i\frac{\varphi''(\omega - \omega_0)^2}{2}-i\frac{\varphi'''(\omega-\omega_0)^3}{6}\right]}
 \end{equation}
The curvature of the zero-loss band is compensated in the second side-band by a light pulse with the same GDD as previously and TOD $\varphi''' = -\beta/2=128\,\mathrm{fs^3}$, as shown in Fig. \ref{fig:03}(b). Further propagation leads to better compression to 11 fs FWHM Fig. \ref{fig:03}(d).\\

\textit{Short Pulse Gating---}Until now, we have focused on modulation using long optical pulses that fully encompass the electron pulse. However, if the electron current is sufficiently high, it becomes possible to use optical pulses that are several times shorter than the electron pulse. This method requires shorter propagation distances for compression and is compatible with electron pulses of higher central energies, that would require distances of several meters to compress with the previous sideband generation method. Additionally, it allows the final compressed sideband duration to be reduced below the photoemission duration, making the approach suitable even for pulses with longer photoemission times.

\begin{figure}
    \centering
    \includegraphics[width=\linewidth]{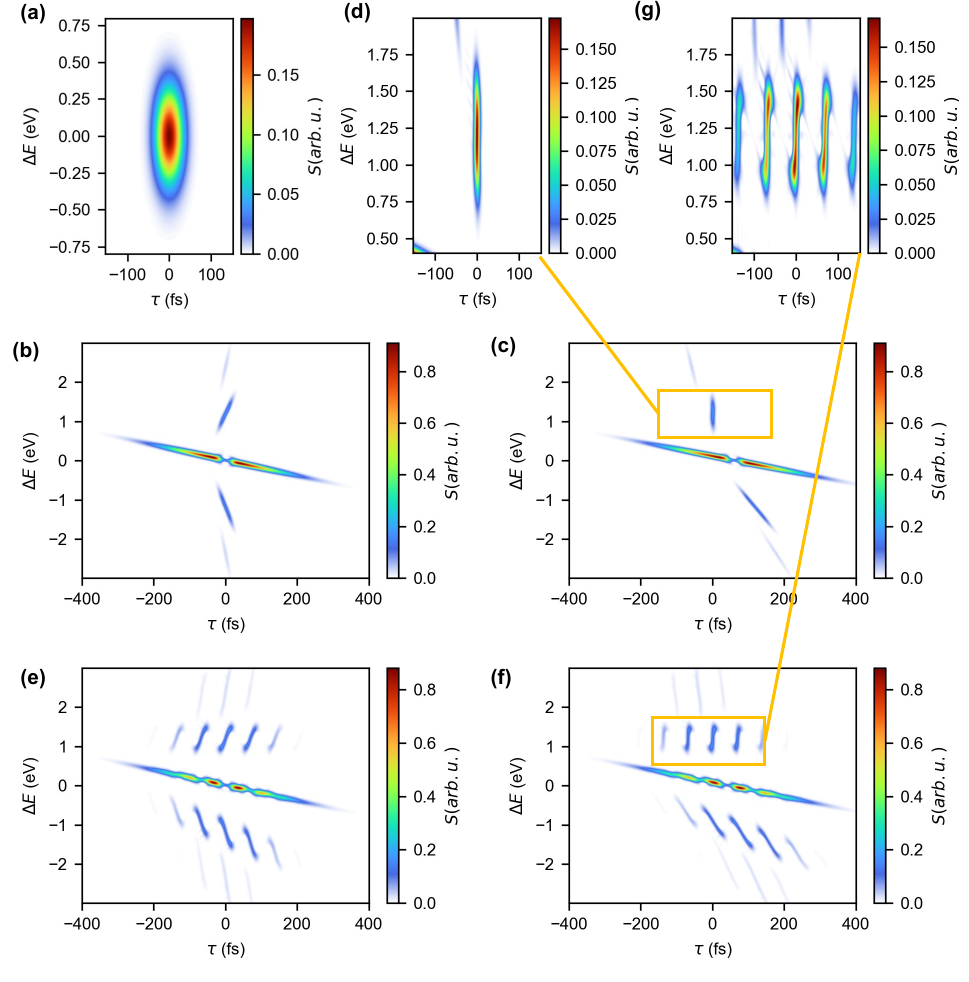}
   
    \caption{(a) Spectrogram of the long initial pulse with FWHM duration 50 fs, (b) of the chirped electron pulse after interaction with a short optical pulse, (c) of the electron pulse state after propagation in free space for $\approx$ 32 cm, (d) close up of the compressed sideband with FWHM temporal duration of $\approx$ 10 fs. Spectrograms (e) of the electron pulse state after interaction with phase- and amplitude-modulated optical field, (f) after propagation for 21 cm, (g) close up of the compressed pulse train.
    } 
    \label{fig:04}
\end{figure}

We model the short pulse gating on an electron wavepacket accelerated to central kinetic energy of $30\,\mathrm{keV}$ ($v_0 = 0.3c$). The energy bandwidth $\delta E$ remains unchanged. The initial duration of the electron pulse is increased to $\tau_0 = 50\,\mathrm{fs}$ [Fig. \ref{fig:04}(a)] and the duration after acceleration and dispersive propagation remains $\tau_\text{el} = 250\,\mathrm{fs}$ FWHM. The GDD coefficient is $\alpha =321 \,\mathrm{fs^2}$, and the TDG coefficient is $\beta = 0$ for simplicity.

While maintaining a constant spectral width of the optical pulse of $0.5 \,\mathrm{eV}$, we set the GDD coefficient to $\varphi''=-\alpha/n$, where $n=11$. After the interaction a short, steeply chirped sideband centered around $1.2\,\mathrm{eV}$ is produced [Fig. \ref{fig:04}(b)]. To optimize the population of the first side-band we set $g=1$.

The propagation distance required for the compression of the first sideband is then reduced to $z_{\text{com}} \approx \frac{m_e v_{\hbar\omega_0}^2 \alpha}{\hbar(n-1)}$, with $v_{\hbar\omega_0}=v(E_0 +\hbar\omega_0)$, while the time duration of the pulse after compression is reduced down to $\tau_{\text{com}} \approx \tau_{\text{0}}/(n-1)$. Specifically for $n=11$, where  $4\%$ of the electrons from the initial pulse are transferred to the first sideband, the compression distance is $z_{\text{com}} \approx 32\,\mathrm{cm}$ and the duration of the compressed electron pulse is $\tau_{\text{com}}\approx 6\,\mathrm{fs}$ [Fig. \ref{fig:04}(c,d)].\\

\textit{Periodic Gating---}The proposed scheme can be used to produce tailored shape in the sideband, that will compress into a train of few-fs pulses in the compression point. While a single few-femtosecond laser pulse generates a single ultrashort electron sub-pulse, a longer optical pulse with periodic modulation in phase and amplitude can produce a train of ultrashort electron sub-pulses.

We start with the same electron pulse parameters as in the previous subsection. Regarding the laser field, periodic frequency modulation is applied
\begin{equation}
\label{freq}
    \omega(\tau) = \omega_{0}\left[1 + m\cos{\left(\Omega \tau\right)}\right] + 2a \tau,
 \end{equation}
where $m=0.25$ is the modulation depth and $\Omega=\omega_0/20$ is the modulation frequency, which needs to be a few times smaller than $\omega_{0}$ for this application. The linear chirp coefficient $a$ is present to compensate for the zero-loss band chirp.

If an optical field with cosine frequency modulation is used, the spacing between the sidebands is also periodically modulated. The optical field in the frequency domain can be written as
\begin{equation}
\begin{split}
    \tilde{\mathcal{E}}(\omega)\propto \sum_{k=-1}^{k=1} \sum_{n=-\infty}^{\infty} (1/2)^{|k|}i^nJ_n \left(\frac{m\omega_0}{\Omega}\right)\times \\  \exp\left[-\frac{(\omega - \omega_{0,kn})^2}{2\sigma^2_\omega} - i\frac{\varphi''(\omega - \omega_{0,kn})^2}{2}\right] ,
\end{split}
 \end{equation}
where $\omega_{0,kn}=\omega_0 - (k+n)\Omega$ and $J_n$ is the Bessel function of the first kind.

Ideally, only the parts of the sideband which have negative chirp must be populated. This is achieved by amplitude modulation of the optical field $\sqrt{I(\tau)}\propto \left[1+\cos{\left(\Omega \tau\right)}\right]/2 $. This ensures that the light field is strong enough for a side-band to be generated only within the selected half-periods [Fig. \ref{fig:04}(e)]. The structured side-band contains around $10\%$ of the initial electron distribution. The negatively chirped sub-pulses are separated in time by $1/2\pi\Omega$, with chirp coefficients $\mathrm{d}E/\mathrm{d}\tau = -\hbar m\omega_{0}\Omega$, and tunable compression distance $z_{\text{com}} \approx \frac{m_e v^2 \alpha} {\hbar(m\omega_{0}\Omega/2a - 1) } $.
 
After propagation for $21\,\mathrm{cm}$, a train of femtosecond pulses compressed several times below the initial duration of the pulse ($\tau_{\text{com}} \approx 6\,\mathrm{fs}$) is generated [Fig. \ref{fig:04}(f,g)]. Considering realistic limitations, such as the finite bandwidth and the finite resolution of a standard optical pulse shaper, this application is still feasible, although minor distortions are to be expected. The temporal envelope of the side-band is expected to be shortened and the sideband modulation depth might be reduced (see Supplemental Material for details \cite{SM}).\\
 
\textit{Discussion---}State-of-the-art light shapers enable precise control over the spectral phase of laser pulses. Our method requires broadband optical pulses with an energy bandwidth of about 0.5 eV to allow straightforward shaping through the introduction of  GDD and TOD. More complex pulse shaping—such as that needed for generating periodic pulse trains—demands a light shaper with both high spectral resolution and sufficient bandwidth. For instance, a shaper with 100 pixels and a bandwidth of 0.7 eV centered around 1.2 eV is already expected to produce a pulse sufficient for the application. For further details, see Supplementary Material, Section II. \cite{SM}.

We need to consider the phase-matching conditions of the chosen light-electron interaction over the broad spectrum of the optical pulse. In the presented simulation results, we approximate the coupling parameter $g(\omega)$ as constant over the bandwidth of the light pulse (see derivation in SM, Section I. \cite{SM}). In reality, the spectral dependence of $g(\omega)$ must be taken into account, and it may be necessary to adjust $I(\omega)$ to achieve the desired temporal probability profile of the generated sidebands. Furthermore, we do not address the effects of transverse momentum transfer during the interaction, which are expected to introduce additional nonlinear chirp due to the unequal trajectory lengths associated with different energies within the sidebands. However, we note that this effect can be mitigated by iteratively shaping the optical field to also compensate for electron delays arising from transverse effects.

Additional temporal broadening can arise from the fact that, in a focused beam, the length of the electron trajectory depends on the propagation angle. This effect can be compensated for by spatio-temporal shaping of the envelope of the driving laser pulse, e.g., by introducing a spherical aberration to the beam leading to a curvature of the envelope in the focal plane.\\

\textit{Conclusion---}We theoretically demonstrate a technique for versatile free-electron manipulation through quantum-coherent interaction with intensity- and phase-shaped optical fields. This approach effectively transfers the state of the art of light pulse shaping techniques to the pulsed electron case, enabling precise control over the electron time-energy characteristics—including spectral width, temporal duration, and both second and higher-order dispersion—in a broadly applicable manner. We demonstrate that the quantum optical electron pulse shaper may achieve electron pulse compression of electrons to few-femtosecond pulse durations without additional spectral broadening of electrons, making this approach ideally suited for ultrafast experiments requiring high time, spatial and spectral resolutions.\\

\section*{Acknowledgments}
The authors acknowledge funding from the Czech Science Foundation (project 22-13001K), Charles University (SVV-2023-260720, PRIMUS/19/SCI/05, GAUK 90424) and the European Union (ERC, eWaveShaper, 101039339). This work was supported by TERAFIT project No. \text{CZ}.02.01.01/00/22\_008/0004594 funded by OP JAK, call Excellent Research.\\

\textit{Data Availability---} The data supporting the findings of this study are openly available at \cite{Data}.

\textit{Code Availability---} The code used for the numerical simulations is available upon request from the corresponding author.


\bibliography{apssamp}
\nocite{Case2008, Abajo2021, Pi25}

%

%

\clearpage
\onecolumngrid
\setcounter{section}{0}
\setcounter{equation}{0}

\renewcommand{\theequation}{S\arabic{equation}}
\section*{Supplemental Material}
\title{Supplemental Material: Quantum Optical Electron Pulse Shaper}

\author{Neli Laštovičková Streshkova}
\email{neli.streshkova@matfyz.cuni.cz}
\affiliation{Department of Chemical
Physics and Optics, Faculty of Mathematics and Physics, Charles University, Ke Karlovu 3, Prague CZ-12116, Czech Republic.}
\author{Martin Kozák}%
\email{m.kozak@matfyz.cuni.cz}
\affiliation{Charles University, Faculty of Mathematics and Physics, Ke Karlovu 3, 121 16 Prague 2}

\date{\today}

\maketitle

\section{Theoretical considerations}
The semiclassical non-relativistic Hamiltonian of a charged particle interacting with an electromagnetic field with vector potential $\mathbf{A}$ is:
\begin{equation}
\label{eqn:s01}
    \hat{H} = \frac{1}{2m_{e}}\left( \hat{\mathbf{p}}+e\mathbf{A}\right)^{2}=\hat{H}_{0}+\hat{H}_{\text{int}},
 \end{equation}
where $e>0$ is the elementary charge, $m_e$ is the electron mass and $\hat{\mathbf{p}}$ is the electron momentum operator. The free-particle Hamiltonian $\hat{H}_0=\hat{\textbf{p}}^2/2m_{e}$ describes the evolution of the wavepacket in free space, while the interaction Hamiltonian $\hat{H}_{\text{int}}$ describes the interaction with an electromagnetic field.

\subsection{Before interaction}

We describe the electron state in time-energy space, which is fully equivalent to the conventional coordinate-momentum representation. For a narrow spectral width, the relativistic energy-momentum relationship is approximately linear: $ E \approx E_0 + v_0(p-p_0)$, where $E_0$ and $v_0$ are the relativistic energy and group velocity at the central momentum $p_0$, respectively. 
The temporal profile of the pulse is captured in the propagation-shifted time $ \tau = t - z/v_0$, assuming that the electrons are propagating in the $z$ direction.

Before the interaction the pure state of the electron pulse is expressed by the wavefunction in the energy domain:
\begin{equation}
\label{eqn:s02}
    \tilde \psi(E) = \mathcal{N}_E e^{ \left[-\frac{(E-E_0)^2}{2\sigma_E^2}\right]}e^{ \left[-i\frac{\alpha (E-E_0)^2}{2\hbar^2}\right]}e^{ \left[-i\frac{\beta (E-E_0)^3} {6 \hbar^3}\right]},
 \end{equation}
where $\mathcal{N}_E$ is a normalization constant, $\sigma_E = \delta E/\sqrt{4\ln(2)}$ the energy width, $\alpha$ is the GDD coefficient and $\beta$ is the TOD coefficient. The time-domain representation is connected to the energy representation via Fourier transform $\psi(\tau)=\mathcal{F}\left\{\tilde{\psi}(E)\right\}$.

To account for a partially coherent state, we introduce the density matrix $\rho(\tau,\tau^*)$
\begin{equation}
\label{eqn:s03}
    \rho(\tau,\tau^*) = \sum_{\tau_0} w_{\tau_0} \psi(\tau - \tau_0) \psi^{\dagger}(\tau^*-\tau_0),
 \end{equation}
 where $\tau^*$ is the second time coordinate of the matrix, $w_{\tau_0}$ is the probability of emission at time $\tau_0$ given by the expression
\begin{equation}
\label{eqn:s04}
    w_{\tau_0} = \mathcal{N}_{w}\exp{ \left[ -\frac{8\ln{2}\tau_0^2}{\tau_{\text{pe}}^2} \right]},
 \end{equation}
where $\mathcal{N}_{w}$ is a normalization constant,  $\tau_{\text{pe}}$ is the FWHM duration of the photoemission optical pulse intensity envelope. Additionally, in the time representation of the wavefunction we only consider the envelope function of the plane wave, omitting the fast oscillation with frequency $E_0/\hbar$.

In phase space we represent the electron state via its Wigner function $W(E,\tau)$, derived from the density matrix $\rho(\tau,\tau^*)$ via Wigner-Weyl transformation \cite{Case2008}. To model a dispersion-free state, we set $\alpha = \beta = 0$ in Eq. (\ref{eqn:s02}), compute the partially incoherent density matrix $\rho_0(t,t^*)$ via Eq. (\ref{eqn:s03})  and apply the Wigner transformation to obtain $W_0(E,t)$. This state has a probability density envelope FWHM duration of $\tau_0 = (\tau_{\text{coh}}^2 + \tau_{\text{pe}}^2)^{1/2}$. For a dispersed state, we use nonzero $\alpha$ and $\beta$ to get $\rho_1(\tau,\tau^*)$ and $W_1(E,\tau)$ with an extended FWHM duration $\tau_{\text{el}} = (\tau_{0}^2 + \tau^2_{\text{chirp}})^{1/2}$, where $\tau_{\text{chirp}}$ accounts for dispersion induced elongation due to propagation between the photocathode and the interaction plane.

\subsection{Interaction}

The interaction with light occurs on much shorter time scales than the free space propagation, therefore it can be treated separately. In the non-recoil approximation, we calculate the effect of the interaction as a perturbation induced phase given by the integral of the Hamiltonian over the classical electron trajectories, described by the interaction time $\theta$ \cite{Abajo2021, Feist2020, Priebe2017, Vanacore2018a}
\begin{equation}
\label{eqn:s05}
    \phi(\textbf{r},t) = \left[-\frac{1}{\hbar}\int^{t}_{-\infty} \hat{H}_{\text{int}}(\textbf{r}+\textbf{v}_0(\theta-t),\theta)\,\mathrm{d}\theta\right],
 \end{equation}
where $\mathbf{v}_0$ is the electron pulse group velocity vector. We define the evolution operator $\hat{U}(\textbf{r},t) = e^{i\phi(\textbf{r},t)}$. Applying the Coulomb gauge and the non-recoil approximation the interaction Hamiltonian becomes $H_\text{int}(\mathbf{r},t)=e\mathbf{p_0}\cdot\textbf{A}(\mathbf{r},t)/m_e$, where the ponderomotive term is neglected.

For a monochromatic electric field with frequency $\omega$, the vector potential is expressed in terms of the electric intensity $\mathbf{A}(\mathbf{r},t) = (-ic/\omega)\mathbf{E}(\mathbf{r},t) + c.c.$. A polychromatic optical incident field with spectrum $\tilde{\mathcal{E}}(\omega)=\sqrt{I(\omega)}e^{i\varphi(\omega)}$, where $I(\omega)$ is the spectral intensity and $\varphi(\omega)$ is the spectral phase, can be expressed as a superposition of the monochromatic components
\begin{equation}
\label{eqn:s06}
    \textbf{E}_i(\mathbf{r}, t) = \int_0^\infty \textrm{d}\omega\,\sqrt{I(\omega)}e^{i\varphi(\omega)}e^{-i\omega( t+ \mathbf{s_i}\cdot \mathbf{r}/c)},
 \end{equation}
where $\mathbf{s}_i$ is the unit vector in the direction of propagation. In the time domain and space we can also express the incident electric field in terms of intensity envelope $I(\mathbf{r},t)$ and phase $\varphi(\mathbf{r},t)$
\begin{equation}
\label{eqn:s07}
    \textbf{E}_i(\mathbf{r}, t) \propto \sqrt{I(\mathbf{r},t)} \exp{\left[-i\varphi(\mathbf{r},t)\right]}.
 \end{equation}
Expanding the phase into polynomial series of the wave argument $(t - \mathbf{s}_i\cdot \mathbf{r}/c)$ we obtain $\varphi(\textbf{r},t) = \omega_0(t-\mathbf{s}_i\cdot \mathbf{r}/c) + a(t-\mathbf{s}_i\cdot \mathbf{r}/c)^2 + ...$, where $\omega_0$ is the central angular frequency and $a$ is the linear chirp coefficient.

In the presence of a membrane or a nanostructure, the total electromagnetic field is 
\begin{equation}
\label{eqn:s08}
    \textbf{E}(\mathbf{r}, t) = \int_0^\infty \textrm{d}\omega\,\sqrt{I(\omega)}e^{i\varphi(\omega)}\mathbf{f}_0(\mathbf{r},\omega)e^{-i\omega t},
 \end{equation}
where $\mathbf{f}_0(\mathbf{r},\omega)$ contains the spatial dependence of the field component on frequency $\omega$ \cite{Morimoto2018a, Vanacore2018a}.

For further evaluation, we consider 1D propagation of the electrons along $z$ with momentum vector $\mathbf{p}_0=(0,0,p_0)$, group velocity vector $\mathbf{v}_0=(0,0,v_0)$ and we only take into account the $z$ component of the electric field $f_{0z}$. Plugging Eq. (\ref{eqn:s08}) into Eq. (\ref{eqn:s05}), we evaluate the evolution operator:
\begin{equation}
    U(z,t) = \exp{\left\{-i2 \text{Im}\int_0^\infty \mathrm{d}\omega
\, g(\omega)\sqrt{I(\omega)}
    e^ {i \varphi(\omega) }e^{-i\omega(t-z/v_0)} \right\}},
\label{eqn:s09}
 \end{equation}
where $g(\omega)$ is the interaction coupling strength for a given frequency
\begin{equation}
\label{eqn:s10}
    g(x,y,\omega) = \frac{e}{\hbar\omega} \int_{-\infty}^{z}\mathrm{d}z'\,f_{0z}(x,y,z';\omega)e^{i\omega z' /v_0}.
 \end{equation}
Assuming that the coupling does not change significantly within the optical pulse bandwidth, we set $g(\omega)\approx g$. We transform to the shifted time coordinate $\tau = t-z/v_0$ and finally we obtain
\begin{equation}
\label{eqn:s11}
    U(\tau) = \exp{\left\{-i2 |g|\sqrt{I(\tau)}
    \sin{[\varphi(\tau)+\arg{g}]} \right\}}.
 \end{equation}
For convenience, we shift all the factors to $g$, so that $I(\tau)$ is normalized to 1 at the maximum. The density matrix describing the state of the electrons after the interaction is:
\begin{equation}
\label{eqn:s12}
    \rho_2(\tau,\tau^*) = U(\tau)\rho_1(\tau,\tau^*)U^{\dagger}(\tau^*),
 \end{equation}
and the corresponding Wigner function $W_2(E,t)$ is obtained via Wigner transformation.

\subsection{After interaction}

As a result of the interaction, sidebands with shaped time-dependent energy evolution are generated in the Wigner function $W_2(E,\tau)$, which are expected to be altered by dispersive propagation in vacuum for time  $\theta_{\text{prop}}$. We emphasize that $\tau$ is the detection time within the electron pulse, whereas $\theta$ is the running evolution time.  The evolution is described as coordinate transformation governed by the free-electron Hamiltonian $\hat{H}_0$ with the Hamilton equation of motion derived from its classical analogue $H_0$. The new coordinates after propagation time $\theta_{\text{prop}}$ are 
\begin{align}
   & \tau'-\tau'_0 = \tau - \tau_0 + \frac{[v(E)-v_0]\theta_{\text{prop}}}{v_0}, \label{eqn:s13} \\
    &E'-E_0' = E - E_0, \label{eqn:s14}
\end{align}
where $v[E(p)]=\partial E(p)/\partial p$ is the momentum-dependent group velocity.
The transformed Wigner function is $W_3(E,\tau,\theta_{\text{prop}}) = W_2(E',\tau')$.

\section{Feasibility considerations}

Since the required optical phase and amplitude modulation would be achieved by an optical modulator with finite resolution and from a finite bandwidth, we investigate the behavior of the pulse shaper under realistic conditions.

Under idealized conditions, the frequency spectrum of the optical field is continuous and infinitely broad. However, in reality, both the bandwidth and the spectral resolution of optical pulse shapers are finite and will limit the practical implementation and performance of the proposed electron pulse shaper. We consider a finite sampling of $N=100$ pixels, spanning a finite spectral window, typically more narrow than that of the optical pulse. In our example, we use a rectangular window with the width of $\Delta \omega=$ 0.7 eV, centered at 1.2 eV, which is experimentally achievable \cite{Pi25}. 

We discuss the cases - the optical pulse with GDD and TOD dispersion and the optical pulse with periodic amplitude and phase modulation combined with GDD. The respective ideal spectral representations are:

\begin{equation}
\label{eqn:s15}
    \tilde{\mathcal{E}}(\omega)\propto
    \exp\left[-\frac{(\omega - \omega_0)^2}{2\sigma^2_\omega} - i\frac{\varphi''(\omega - \omega_0)^2}{2}-i\frac{\varphi'''(\omega-\omega_0)^3}{6}\right],
\end{equation}

\begin{equation}
\label{eqn:s16}
    \tilde{\mathcal{E}}(\omega)\propto\sum_{k=-1}^{k=1} \sum_{n=-\infty}^{\infty} (1/2)^{|k|}i^nJ_n \left(\frac{m\omega_0}{\Omega}\right)\times \exp\left[-\frac{(\omega - \omega_{0,kn})^2}{2\sigma^2_\omega} - i\frac{\varphi''(\omega - \omega_{0,kn})^2}{2}\right].
\end{equation}

We show the comparison between the interaction of the electron pulse with a light field with an ideal, continuously modulated spectral intensity and a light field, which models a realistic optical pulse that is shaped by a light modulator.

To model the realistic light pulse, we first limit the spectral width by a rectangular window, which is non-zero for the interval $(\omega_0 - \Delta\omega/2; \omega_0 + \Delta\omega/2 )$. Then we divide this interval by the number of pixels $N$ and assign a discrete amplitude and phase value to each of the frequencies, effectively truncating and down-sampling the light pulse frequency spectrum and spectral phase. Formally, the continuous integration in Eq. (\ref{eqn:s06}) is replaced by the discrete sum
\begin{equation}
\label{eqn:08}
    \textbf{E}_i(\mathbf{r}, t) = \frac{\Delta\omega}{N}\sum_{n=1}^{N} \tilde{\mathcal{E}}(\omega_n)e^{-i\omega_n( t+ \mathbf{s_i}\cdot \mathbf{r}/c)},
\end{equation}
where $\omega_n = \omega_0-\Delta\omega (1/2 - n/N))$.

In the case of GDD and TOD correction, the comparison of the ideal and realistic intensity spectra is shown in Fig. S1. Large part of the spectrum around the central frequency remains practically unchanged. Because the central most intense part of the spectrum is also the part that mainly contributes to the generation of the second sideband, the result of the interaction with the realistically shaped optical field is almost identical to the one with ideally shaped optical field. 

On the other hand, in the case of the periodic frequency and intensity modulation of the optical pulse, the spectrum is more complicated, consisting of many peaks (see Fig. S3, blue line). While the discretization to 100 pixels is not detrimental to the peak resolution, important features of the spectrum are not transmitted through the spectral window, resulting in changes in the spectrum of the optical pulse Fig. S3, orange line. After interaction with the realistic optical pulse intensity spectrum, the sub-pulses have narrower spectrum [see Fig. S4]. However, our data demonstrate that downsampling of the spectral intensity and phase of the optical pulse shaper are not detrimental for the operation of the proposed electron pulse shaper.

\begin{figure}
    \centering
    \includegraphics[width=0.5\linewidth]{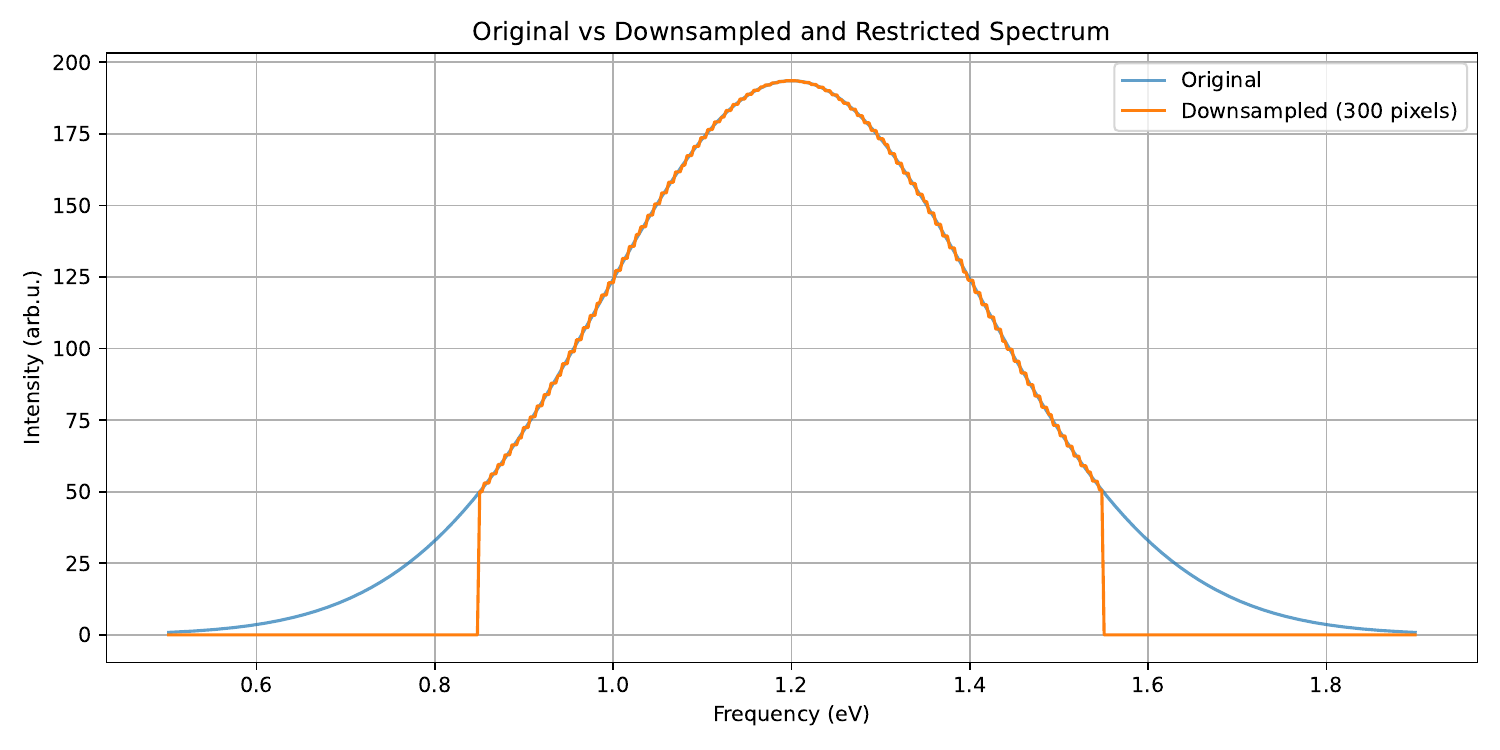}
    \caption{Comparison of the ideal and the down-sampled and frequency-restricted spectrum of light pulse with GDD and TOD.}
    \label{fig:S1}
\end{figure}

\begin{figure}
    \centering
    \includegraphics[width=0.5\linewidth]{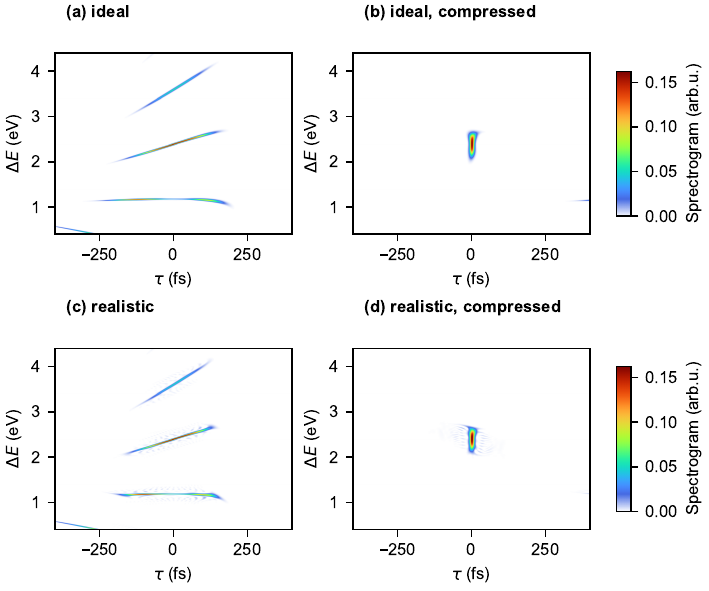}
    \caption{Wigner function of the electron pulse after interaction with the (a) ideal optical field with GDD and TOD, (b) after compression of the second side-band, (c) after itneraction with realistically shaped optical field and after compression (d).}
    \label{fig:S2}
\end{figure}

\begin{figure}
    \centering
    \includegraphics[width=0.5\linewidth]{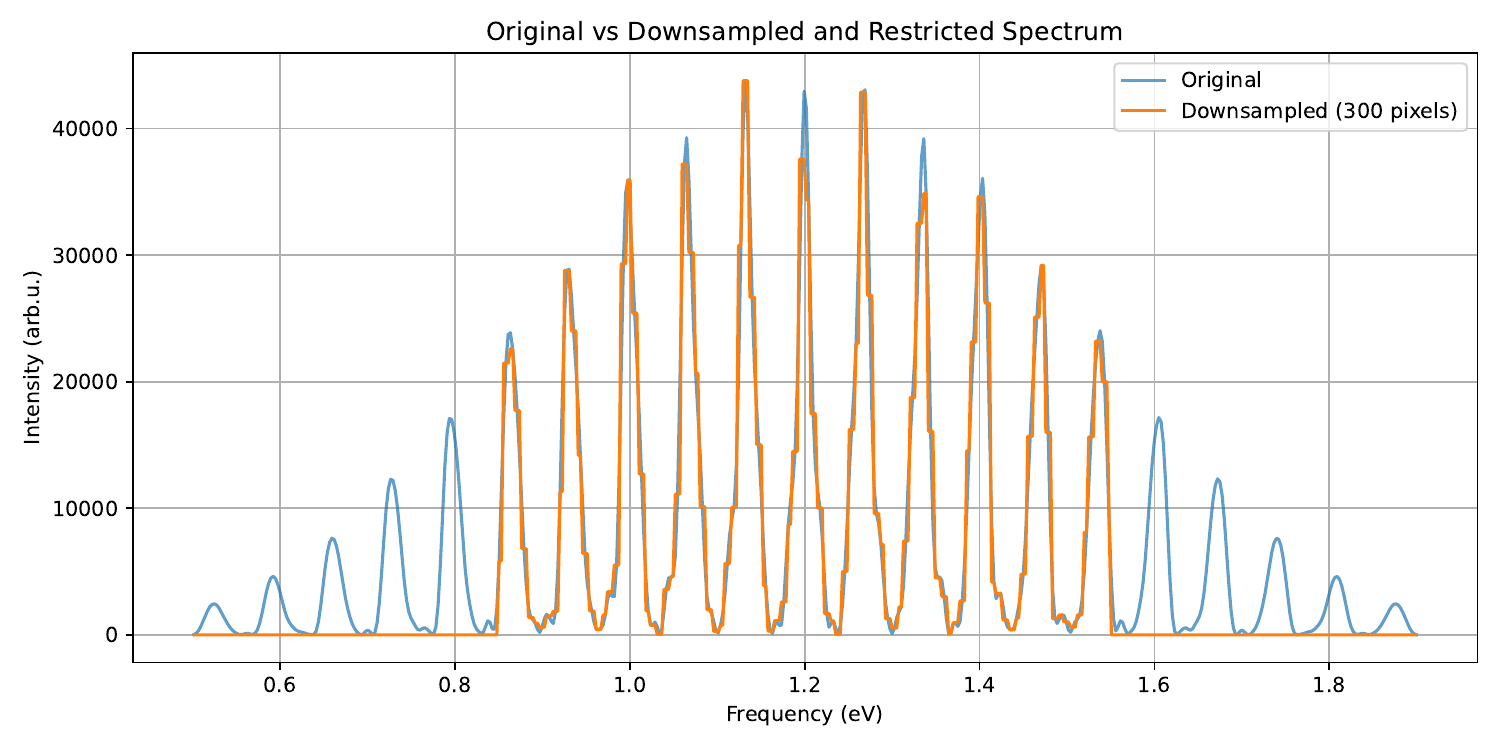}
    \caption{Comparison of the ideal and the down-sampled and frequency-restricted spectrum of light pulse with periodic phase and amplitude modulation.}
    \label{fig:S3}
\end{figure}

\begin{figure}
    \centering
    \includegraphics[width=0.5\linewidth]{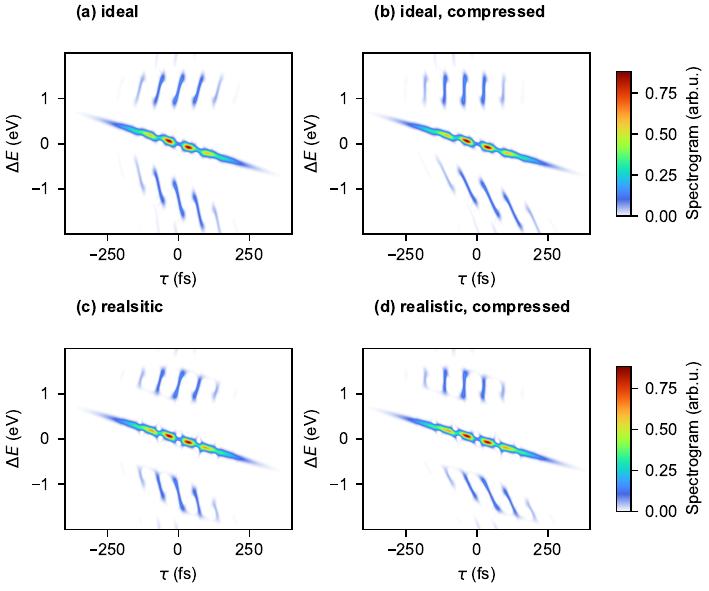}
    \caption{Wigner function of the electron pulse interaction with the ideal optical field with periodic phase and frequency modulation, (a) before and (b) after compression of the first side-band, after interaction with realistically shaped optical field (c) before and (d) after compression.}
    \label{fig:S4}
\end{figure}


\end{document}